\documentclass{elsarticle}

\usepackage[utf8]{inputenc}
\usepackage{hyperref}
\usepackage{amssymb}
\usepackage{float}
\usepackage{epstopdf}
\usepackage{amssymb}
\usepackage{amsthm}
\usepackage{mathtools}
\usepackage{caption}
\usepackage{subcaption}

\DeclarePairedDelimiter\floor{\lfloor}{\rfloor}

\journal{}

\begin{document}
\begin{frontmatter}

\title{Predicting future stock market structure by combining social and financial network information}

\author[London]{Thársis T. P. Souza\corref{cor1}}
\ead{T.Souza@ucl.ac.uk}
\cortext[cor1]{Corresponding author}
\author[London]{Tomaso Aste}
\address[London]{Department of Computer Science, UCL, Gower Street, London, WC1E 6BT, UK}

\begin{abstract}
We demonstrate that future market correlation structure can be predicted with high out-of-sample accuracy using 
a multiplex network approach that combines information from social media and financial data.
Market structure is measured by quantifying the co-movement of asset prices returns, while social structure is 
measured as the co-movement of social media opinion on those same assets.
Predictions are obtained with a simple model that uses link persistence and link formation by triadic closure across both financial and social media layers.
Results demonstrate that the proposed model can predict 
future market structure with up to a 40\% out-of-sample performance improvement compared to a benchmark model that 
assumes a time-invariant financial correlation structure.
Social media information leads to improved models for all settings tested, particularly in the long-term prediction of financial market structure.
Surprisingly, financial market structure exhibited higher predictability than social opinion structure.
\end{abstract}


\end{frontmatter}


\section{Introduction}

Financial markets can be regarded as a complex network in which nodes represent different financial assets and edges 
represent one or many types of relationships among those assets. 
Filtered correlation-based networks have successfully been used in the literature to study financial markets structure particularly 
from observational data derived from empirical financial time series \cite{10.1371/journal.pone.0017994, Mantegna1999, aste2010correlation, Tumminello201040, Tumminello26072005}.
The underlying principle is to use correlations from empirical financial time series to construct a sparse network representing 
the most relevant connections.
Analyses on filtered correlation-based networks for information extraction \cite{1367-2630-12-8-085009,song2008analysis,aste2010correlation} 
have widely been used to 
explain market interconnectedness from high-dimensional data.
Applications include asset allocation \cite{pozzi2013spread}, market stability assessments \cite{morales2012dynamical}, 
hierarchical structure analyses \cite{Mantegna1999, aste2010correlation,Tumminello201040, musmeci2014clustering, song2012hierarchical} 
and the identification of lead-lag relationships \cite{curme2015coupled}.


%
The majority of literature so far has focused on the analysis of financial time series. 
However, in recent years a large amount of information about financial markets has become available from exogenous sources such as social media.   
It is reasonable to conceive that changes in social media sentiment \cite{1507.00955} and changes in asset prices might be related.
Some  previous studies have indeed demonstrated the existence of relationships which in some cases 
indicated that social media can be used to predict changes in asset prices \cite{8252134, 1507.00784, citeulike:13108056, tetlock2007giving, Tobias:2013, mao2014quantifying}. 
When new information hits the markets, investors may react either rationally or irrationally \cite{BondtandThaler1985, shleifer2000inefficient}. 
They may express opinions on social media that can later become market actions, thus enabling opportunities to forecast future asset prices. 
However, it has also been highlighted that not all assets behave in the same way.
Some are more influenced by social media sentiment, while others, on the contrary, 
are more influential on the social media sentiment \cite{2016arXiv160104535S}.
Besides each single financial asset, we address in this study whether the entire stock market structure 
is related to the structure constructed from 
social media sentiment and whether there exist lead-lag relationships exist that can be used for forecasting one structure in terms of the other.
 
We use dynamical Kendall correlations computed over rolling windows to investigate the temporal 
evolution of market structure represented by filtered correlation-based networks constructed from stock market prices and from Twitter sentiment signals.
We generate two networks: one from log-returns of stock prices and the other from Twitter sentiment. 
The two networks are treated as a multilayer problem with two layers of networks that share the same nodes but have different edge sets. 
We investigate whether financial market structure can be better predicted by 
combining past financial information with past social media sentiment information. 
The market structure forecasting problem is formulated as a link prediction problem where we estimate the probability of addition 
or removal of a link in the future based on information about the structure of the financial and social networks in the past.

\section{Methods}

\subsection{Financial and Social Networks}
We selected $N = 100$ of the most capitalized companies that were part of the S\&P500 index 
from 09/05/2012 to 08/25/2017. 
The list of these companies' ticker symbols is reported in the Appendix \ref{sec:comps}.
For each stock $i$ the financial variable was defined as 
the daily stock's log-return $R_i(\tau) = \log{Price(\tau)} - \log{Price(\tau-1)}$, where $Price(\tau)$ designates the closing price at time $\tau$. 
The social media variable was defined as the the social media opinion $O_i$ of stock $i$ 
which was estimated as the total number of bullish daily tweets related to the stock $i$ at time $\tau$.
Twitter sentiment data were provided by PsychSignal.com \cite{PsychSignal}.
In this dataset, a Twitter message was defined to be related to a given stock when its ticker symbol was mentioned. 
The dataset used only English language content and it was agnostic to the country source of the Twitter message.
We have provided further descriptive analytics of the Twitter sentiment dataset used in related literature \cite{8252134, 2016arXiv160104535S}.

Stock returns $R_i$ and social media opinion scores $O_i$ each amounted to a time series of length equals to 1251 trading days.
These series were divided time-wise into $M = 225$ windows $t = 1, 2, \ldots, M$
of width $T = 126$ trading days. 
A window step length parameter of $\delta T = 5$ trading days defined the displacement of the window, i.e., the number of trading days 
between two consecutive windows. The choice of window width $T$ and window step $\delta T$ is arbitrary, and
it is a trade-off between having analysis that is either too dynamic or too smooth. 
The smaller the window width and the larger the window steps, the more dynamic the data are.

To characterize the synchronous time evolution of assets, 
we used equal time Kendall's rank coefficients
between assets $i$ and $j$, defined as
\begin{equation}
 \rho_{i, j}(t) = \sum\limits_{t' < \tau}sgn(V_i(t') - V_i(\tau))sgn(V_j(t') - V_j(\tau)),
\end{equation}
where $t'$ and $\tau$ are time indexes within the window $t$ and $V_i \in \{R_i, O_i\}$.

Kendall's rank coefficients fulfill the condition $-1 \leq \rho_{i, j} \leq 1$ and form the $N \times N$ correlation matrix
$C(t)$ that served as the basis for the networks constructed in this paper. To construct 
the asset-based financial and social networks, we defined a distance between a pair of stocks. This
distance was associated with the edge connecting the stocks, and it reflected the level at which they were correlated.
We used a simple non-linear transformation $d_{i, j}(t) = \sqrt{2(1 - \rho_{i,j}(t))}$ to obtain distances with the
property $2 \geq d_{i,j} \geq 0$, forming a $N \times N$ symmetric distance matrix $D(t)$. 

We extracted the $N(N-1)/2$ distinct distance elements from the upper triangular part of the distance matrix
$D(t)$, which were then sorted in an ascending order to form an ordered sequence 
$d_1(t), d_2(t), \ldots, d_{N(N-1)/2}(t)$. Since we require the graph to be representative of the market,
it is natural to build the network by including only the strongest connections. This is a network filtering procedure 
that has been successfully applied in the construction of \textit{asset graphs} for the analyses of market structure \cite{1402-4896-2003-T106-011, refId0-Onnela-2004}.
The number of edges to include is arbitrary, and we included those from the bottom quartile, 
which represented the 25\% shortest edges in the graph (largest correlations), thus giving $E(t) = \{d_1(t), d_2(t), \ldots, d_{\floor{N/4}}(t)\}$.

We denoted $E^{F}(t)$ and $E^{S}(t)$ as the set of edges constructed from the distance matrices derived from 
stock returns $R(t)$ and social media opinion $O(t)$, respectively.
Two networks were considered as two layers of a duplex structure  
$\mathcal{G} = \{G^{F}, G^{S}\}$ where $G^{F} = ( V, E^{F} )$, $G^{S} = ( V, E^{S} )$  
and $V$ is the vertex set of stocks which is common to both layers.

\subsection{Persistence}
The state of an edge between vertices $u$ and $v$  in the financial layer at time $t$ was represented 
with the corresponding adjacency matrix element $E^F_{u,v}(t)$: a binary variable with $E^F_{u,v}(t)=1$ 
indicating the existence of the edge and $E^F_{u,v}(t)=0$ its absence. 
Analogously, the variable $E^S_{u,v}(t)$  accounted for the presence or absence of edge $(u,v)$ in the social ($S$) layer.
The variable $E_{u,v}(t)=E^F_{u,v}(t) \vee E^S_{u,v}(t)=1$ indicates instead the presence of at least one 
edge between $u$ and $v$ in the two layers; $E_{u,v}(t)=0$ indicates that no edges are present between $u$ and $v$ in any layer.

\subsection{Triadic Closure}
Let $\mathcal{N}_{uv}$ be the set of nodes that are common neighbors to vertices $u$ and $v$.
We defined the triadic closure $T^F_{u,v}(t)$ of an edge $(u,v)$ at layer $F$ and time $t$ as the mean of the clustering coefficients of vertices in  $\mathcal{N}_{uv}$:
\begin{equation}
T^F_{u,v}(t) = \frac{1}{|\mathcal{N}_{uv}|}\sum_{i \in \mathcal{N}_{uv}} C^F_i(t),
\end{equation}
where term $C^F_i$ is the clustering coefficient of node $i$ which accounts for the 
fraction of triads in the neighbors of $i$  that are closed in triangles. This is defined as
\begin{equation}
C^F_i = 2 \frac{\mbox{Number of triangles with a vertex on $i$}}{k_i (k_i - 1)} =  \frac{\sum_{j,k\in \mathcal{N}_i} E^F_{j,k}}{k_i (k_i - 1)}, 
\end{equation}
where $k_i$ is the degree of vertex $i$ and $\mathcal{N}_i$ is the neighborhood of $i$.

In the multiplex case, we kept the same definition but allowed triangles to form across several layers \cite{Battiston2017, 1367-2630-17-7-073029}. 
For the multiplex case, we used the symbol $T_{u,v}(t)$.
\captionsetup[subfigure]{labelformat=empty}
\begin{figure}[!h]
\centering
\begin{subfigure}{.45\textwidth}
\caption{\textbf{A)}}
  \centering
  \includegraphics[width=.9\linewidth]{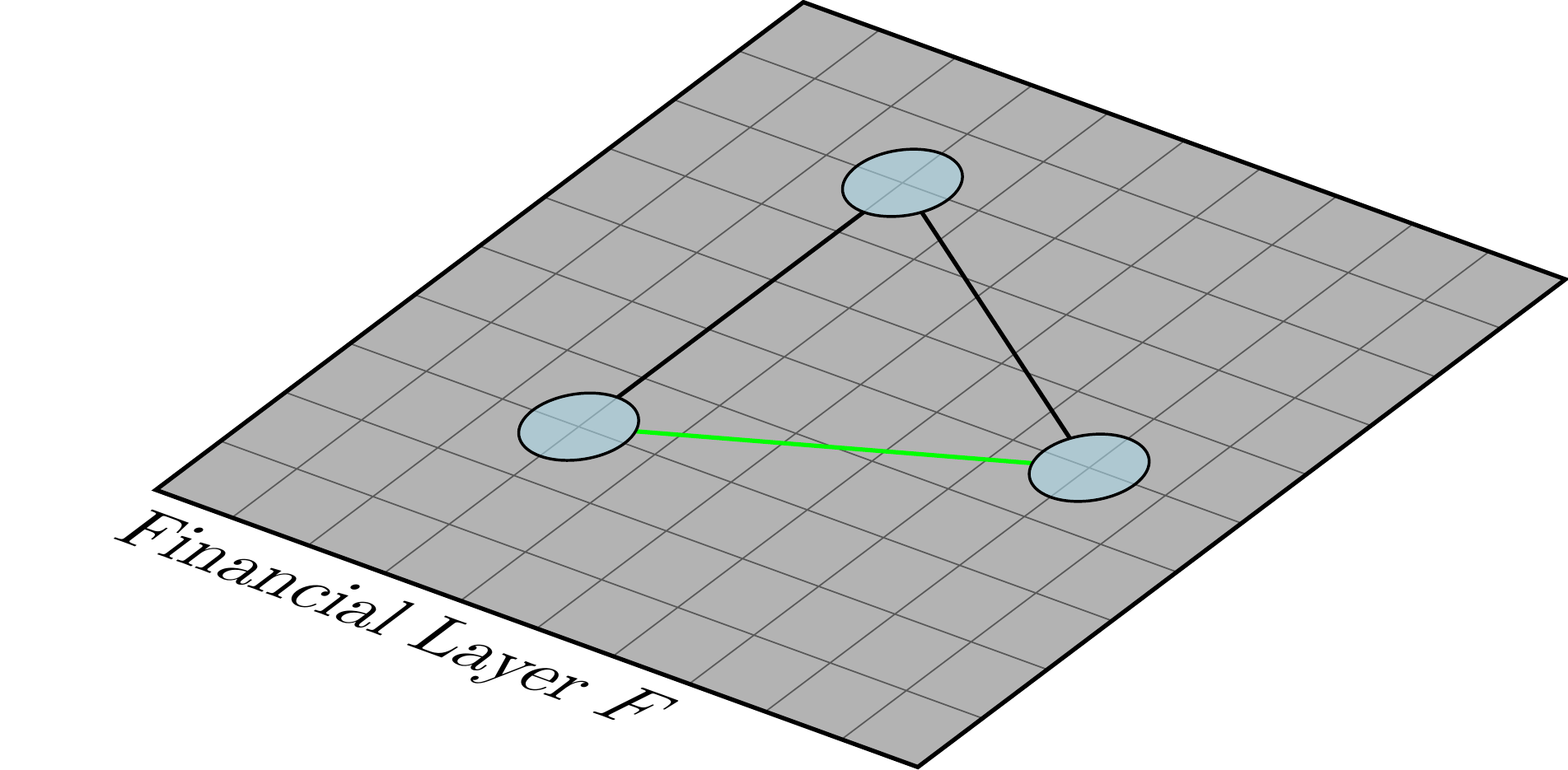}
  \label{fig:sub1}
\end{subfigure}%
\begin{subfigure}{.45\textwidth}
\caption{\textbf{B)}}
  \centering
  \includegraphics[width=.9\linewidth]{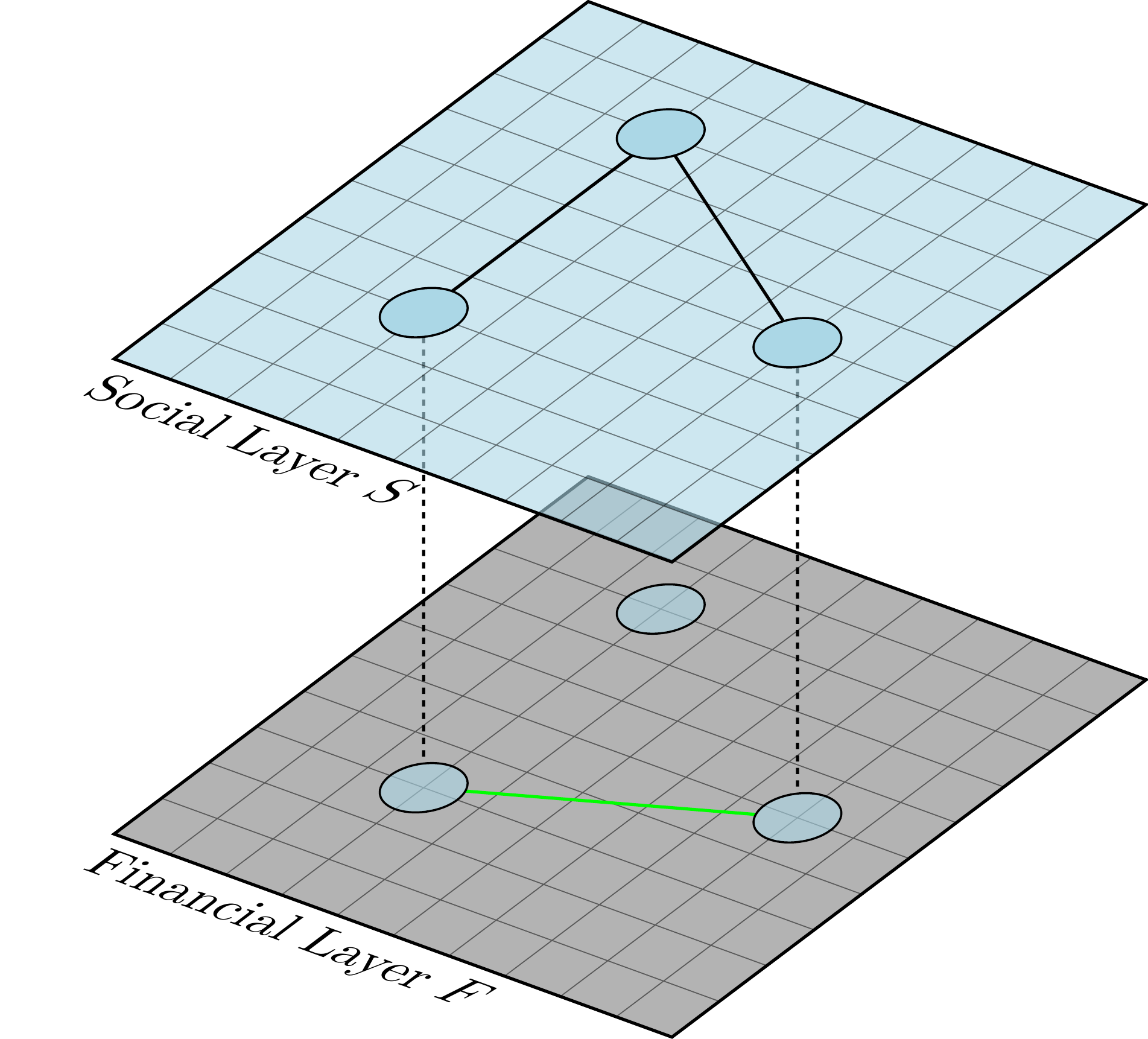}
  \label{fig:sub2}
\end{subfigure}
\caption{Triads on a single layered network (Panel A) and on a multiplex network (Panel B). 
The clustering coefficient of node $i$ accounts for the fraction of triads in the neighborhood of $i$ that are closed in triangles.
The triadic closure of an edge $(u,v)$ at layer $F$ is a function of the clustering coefficients of the common neighbors of the vertices $u$ and $v$.
Triangles can be formed in a single layer or across layers.}
\label{fig:TRIADS}
\end{figure} 
 
\subsection{Link Prediction}
\label{sec:link}
We aim to predict the probability that an edge is inserted or removed in the financial network, $G^{F}(t+h)$, at a future time $t+h$ 
by using the information about the past structures of the financial and social networks at previous times $t' \le t$.  
For this purpose we considered two mechanisms:  
\begin{itemize}
\item[1)] the tendency of an edge present at a previous time to persist in the future ({\it edge persistence}); 
\item[2)] the propensity of triangles within or across layers to close ({\it triadic closure}).
\end{itemize}
The mechanism of growth by triadic closure is based on a principle of transitivity, often observed in real-world networks, 
where there is a tendency to form triangles. Under this principle, two nodes tend to be connected if they share common neighbors with high transitivity, i.e., propensity to close triangles.

The probability that an edge will be inserted in the future is computed by means of a logistic regression of the edge persistence and the triadic closure coefficients. 
We estimated regression coefficients by best fitting on a training set which was composed of rolling windows of 126 trading days that 
initially ranged from 09/05/2012 to 09/10/2014. 
Predictions concerning the presence of edges in the financial network were made at $h=1$ to $h=20$ weeks ahead of the end of the training set.
The test set initially ranged from 09/17/2014 to 08/25/2017.
The procedure was repeated by moving the training window forward in 1-week steps. 

The probability $p_{u,v}(t+h)$ to observe vertices $u$, $v$ connected by an edge at $t+h$ can be inferred
in terms of the set of previous triadic closure coefficients, 
$T_{u,v}(t)$, and edge persistence scores $E_{u,v}(t)$. We first considered a restricted model that used financial information only, which 
is given by the following: 
 \begin{equation}
   \log\frac{p^F_{u,v}(t+h)}{1 - p^F_{u,v}(t+h)} = \tilde\beta_0^h + \tilde\beta^h T^F_{u,v}(t) + \tilde\gamma^h E^F_{u,v}(t).
   \label{eq:logit2}
 \end{equation}
For this restricted model, we performed a 1-step ahead prediction for $h \in (1, 2, \ldots, 19, 20)$ weeks.
 
To calibrate the parameters in Eq. \ref{eq:logit2}, we considered a training window of $W = 126$ days which ends at time $t$.
The log-likelihood function \cite{faraway2006elm} over the training window for the logistic model from Eq. \ref{eq:logit2} is given by
\begin{equation} \label{eq:loglik2}
\begin{split}
 \mathcal{L}^F(t) =& \, \sum\limits_{t'=t-W+1}^{t}\,\sum\limits_{uv \in E^{F}(t'+h)} - \log{(1 + e^{\tilde\beta_0^h + \tilde\beta^h T^F_{u,v}(t') + \tilde\gamma^h E^F_{u,v}(t')})} \, + \\ 
 & \sum\limits_{t'=t-W+1}^{t}\,\sum\limits_{uv \in E^{F}(t'+h)} (1 - E_{uv}^F(t'+h))(\tilde\beta_0^h + \tilde\beta^h T^F_{u,v}(t') + \tilde\gamma^h E^F_{u,v}(t')).
\end{split}
\end{equation}

We differentiated the log-likelihood function given by Eq. \ref{eq:loglik2} in order to find maximum
log-likelihood estimates for the coefficients of Eq. \ref{eq:logit2}.  
  
To verify whether the multiplex information is relevant in the prediction of links in the financial network compared to past a financial network alone, 
we considered a full regression model that takes the set of previous triadic closure coefficients and edge persistence 
from the financial layer ($T^F_{u,v}(t), E^F_{u,v}(t)$), social layer ($T^S_{u,v}(t), E^S_{u,v}(t)$) and the multiplex network ($T^F_{u,v}(t), E^F_{u,v}(t)$).
The full model is  
\begin{equation} \label{eq:logit}
\begin{split}
   \log\frac{p_{u,v}(t+h)}{1 - p_{u,v}(t+h)} =& \, \beta_0^h + \beta_1^h T^F_{u,v}(t) + \beta_2^h E^F_{u,v}(t) \, + \\
   & \gamma_1^h T^S_{u,v}(t) + \gamma_2^h E^S_{u,v}(t) + \theta_1^h T_{u,v}(t) + \theta_2^h E_{u,v}(t).
 \end{split}
 \end{equation}

The log-likelihood function $\mathcal{L}(t)$ of the full model in Eq. \ref{eq:logit} and
the model fitting can be obtained in an analogous manner to the previously performed procedure for the restricted model from Eq. \ref{eq:logit2}.

The likelihood ratio statistic is
 \begin{equation}
\lambda(t) = -2 (\mathcal{L}_{max}(t) - \mathcal{L}_{max}^F(t)),
 \end{equation}
where $\mathcal{L}_{max}(t)$ and $\mathcal{L}_{max}^F(t)$ are,respectively, the maxima of the log-likelihood functions of the full and 
restricted models in the training set window.
The likelihood ratio statistic $\lambda(t)$ can be assumed to follow a $\chi^2$ distribution \cite{faraway2006elm} with 4 degrees of freedom where a value of $\lambda > 18.47$ 
is assumed to be statistically significant at $p = 0.001$.
In that case, there is evidence to accept the full model that considers social and financial information 
over the restricted model that considers financial information only.

The model performance was estimated by counting both the true positives (edges predicted to be there and indeed present in the future network) 
and the false positives (edges predicted to be there but not present in the future network) 
and measuring of AUC (area under the receiver operating characteristic curve) in the test set that originally ranged from 09/17/2014 to 08/25/2017.
AUC ranges from 0.50 to 1.00, with higher values indicating that the model discriminates better between the two categories of edge-present and edge-absent.

\section{Results}

\subsection{Market structure dynamics}
\begin{figure}[!h]
\centering
\scalebox{0.28}{\includegraphics{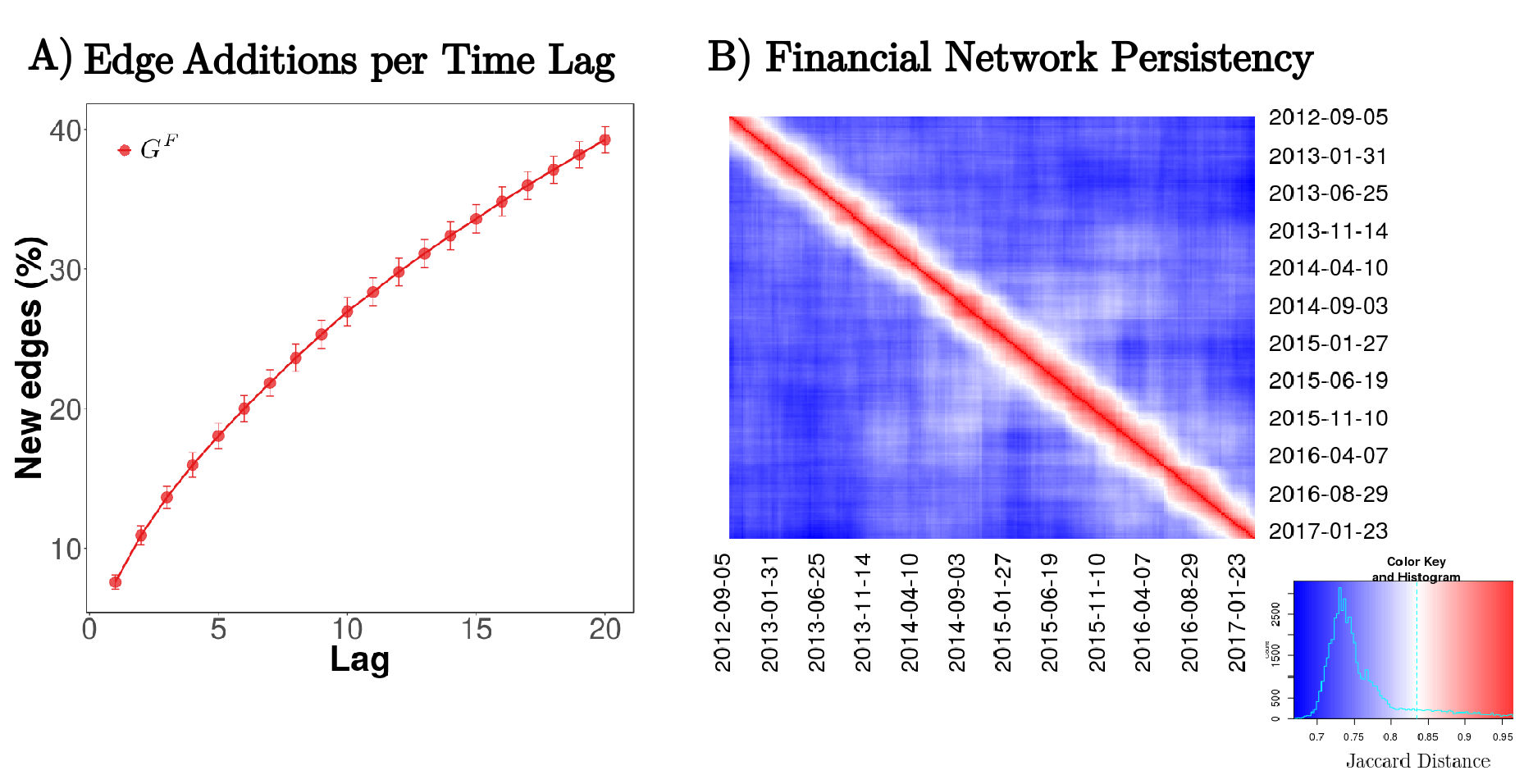}}
\caption{\textbf{Evidence that financial correlation structure changes considerably with time}. 
Panel A) shows the mean percentage of new edges in the financial network at time $t+h$ with respect to the edge set at time $t$  ($1 \leq h \leq 20$ trading weeks). 
We observe that edges change considerably in the financial network with almost 40\% of edges in financial networks changing after a period of $h = 20$ trading weeks. 
Panel B) shows the cross-similarity among financial networks measured as the Jaccard Distance between $G^{F}(t')$ and $G^{F}(t)$ with $t$ and $t'$ 
ranging from 09/05/2012 to 21/02/2017. We observe that edge changes (persistence) are quite stable overtime, i.e., 
the number of edges that change is similar throughout the period. 
Network $G^{F}(t)$ are constructed at each time $t$ from a correlation structure estimated from a sliding window of 126 trading days starting at time $t$.
The windows move with time step of 1 trading week.
Error bars in Panel A) indicate standard error.
}
\label{fig:TRIADS1-rolling}
 \end{figure}

We first investigated financial network persistence by comparing the financial network $G^{F}(t)$ 
at time $t$ with a future financial network, $G^{F}(t+h)$ at $h$ steps ahead.
To quantify the changes in the correlation network structure, we used two measures: 
A) the fraction of new edges in $G^{F}(t+h)$ that were not present in $G^{F}(t)$; 
B) the Jaccard Distance, defined as
$$ Jaccard(G^{F}(t'), G^{F}(t)) = \frac{\|G^{F}(t') \cap G^{F}(t)\|}{\|G^{F}(t') \cup G^{F}(t)\|}.$$
Results are reported in Fig. \ref{fig:TRIADS1-rolling}, panels A) and B), respectively.

Fig. \ref{fig:TRIADS1-rolling} panel A) shows the mean percentage of new edges in the financial network at time $t+h$ with respect to the edge set at time $t$  ($1 \leq h \leq 20$ trading weeks). 
We observe that edges change considerably in the financial network with almost 40\% of edges in financial networks changing after a period of $h = 20$ trading weeks. 
Fig. \ref{fig:TRIADS1-rolling} panel B) shows the cross-similarity among financial networks measured as the Jaccard Distance between $G^{F}(t')$ and $G^{F}(t)$ with $t$ and $t'$ 
ranging from 09/05/2012 to 21/02/2017. We observe that edge changes (persistence) are quite stable overtime, i.e., 
the number of edges that change is similar throughout the period. 
Hence, results indicate that the constructed financial networks are time-variant across the entire period studied, 
with a stable rate of edge changes over time.


\subsection{Prediction of Stock Market Structure} 
\begin{figure}[!h]
\centering
\scalebox{0.36}{\includegraphics{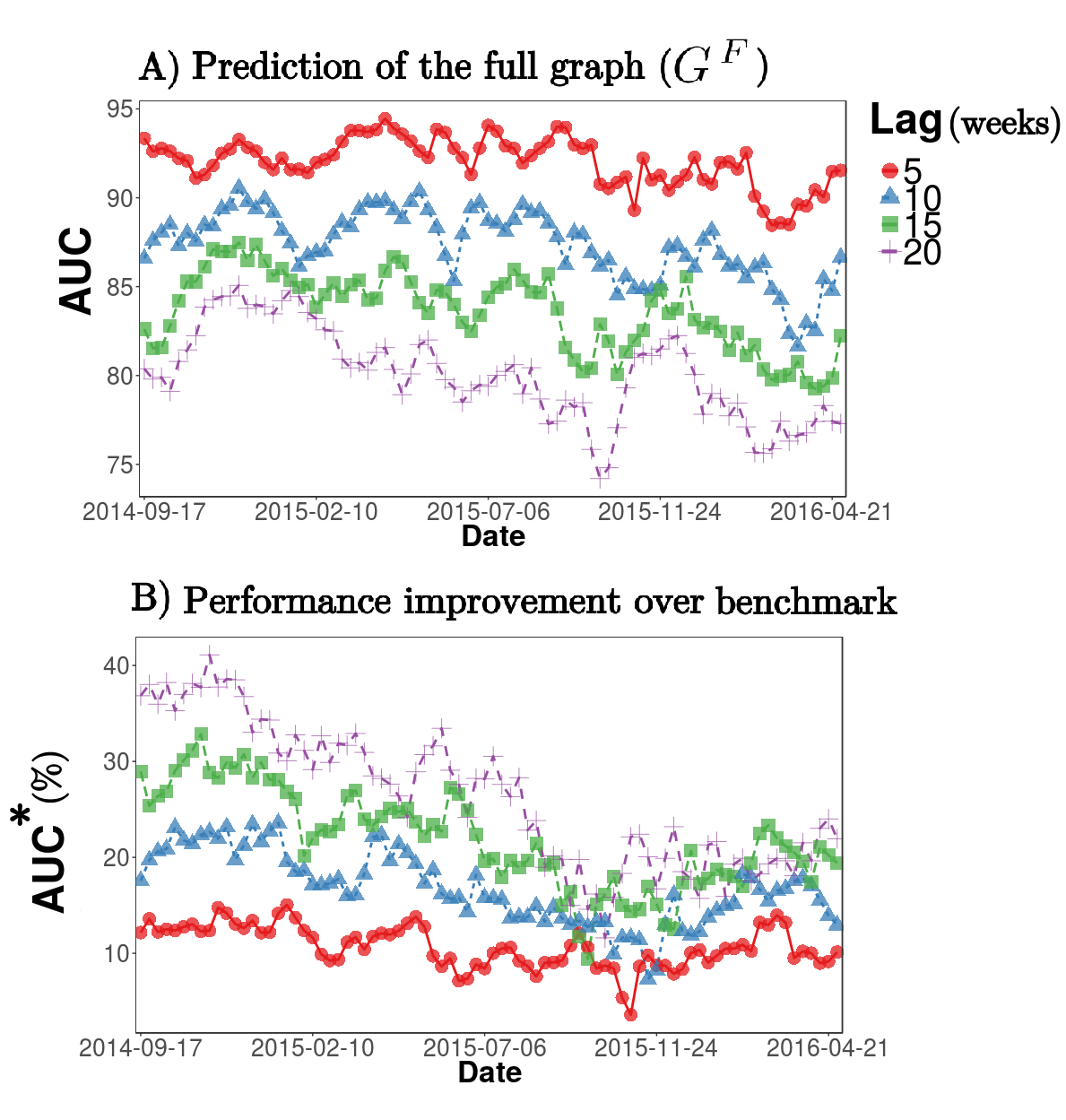}}
\caption{\textbf{Evidence of high out-of-sample performance in financial network link prediction.}
Models were trained in an expanding window with initial start and end dates 09/05/2012.
and 09/10/2014, respectively. Test period ranges from 09/17/2014 and 08/25/2017. 
Plots display the performance results (AUC) of a model to predict edges in a financial network at time $t + h$ trained with
information up to date $t$. 
Panel A) shows the performance obtained in the prediction of out-of-sample edges for $h \in (1, 5, 10, 15, 20)$ trading weeks. 
Panel B) shows the performance improvement ($AUC^*$) compared to a naive benchmark that assumes that the 
correlation structure is time-invariant, i.e., $G^{F}(t+h) = G^{F}(t)$.
}
\label{fig:TRIADS3-rolling}
 \end{figure}

We used Eq. \ref{eq:logit} to predict a the financial network $G^{F}(t+h)$ at a future time $t+h$ 
by using the information about the past structures of the financial and social networks at previous times $t' \le t$.  
Fig. \ref{fig:TRIADS3-rolling} panel A) shows the performance obtained in the prediction of 
out-of-sample edges for $h \in (1, 5, 10, 15, 20)$ trading steps ahead. 
We achieved an overall high out-of-sample performance in financial network link prediction, 
with performances in the range of 73\% to 95\% depending 
on time-lag and time-period. Prediction power improved with a smaller time lag.

We compared our results to those obtained using a benchmark model that assumes that correlation structure is 
time-invariant, i.e., $G^{F}(t+h) = G^{F}(t)$.
The performance improvement against the benchmark is estimated as $AUC^* = (AUC - 0.5)/(\widehat{AUC} - 0.5) - 1$, where $AUC$ 
represents the performance of the proposed model and $\widehat{AUC}$ is the performance of the benchmark.  
From Fig. \ref{fig:TRIADS3-rolling} panel B), we observe that the higher the time lag, the higher the performance improvement over the benchmark.
Let us note that performance improvement over the naive benchmark reached values as high as 40\% for a long-term prediction with a lag of 20 trading weeks.
 
\begin{figure}[!h] 
\centering
\scalebox{0.28}{\includegraphics{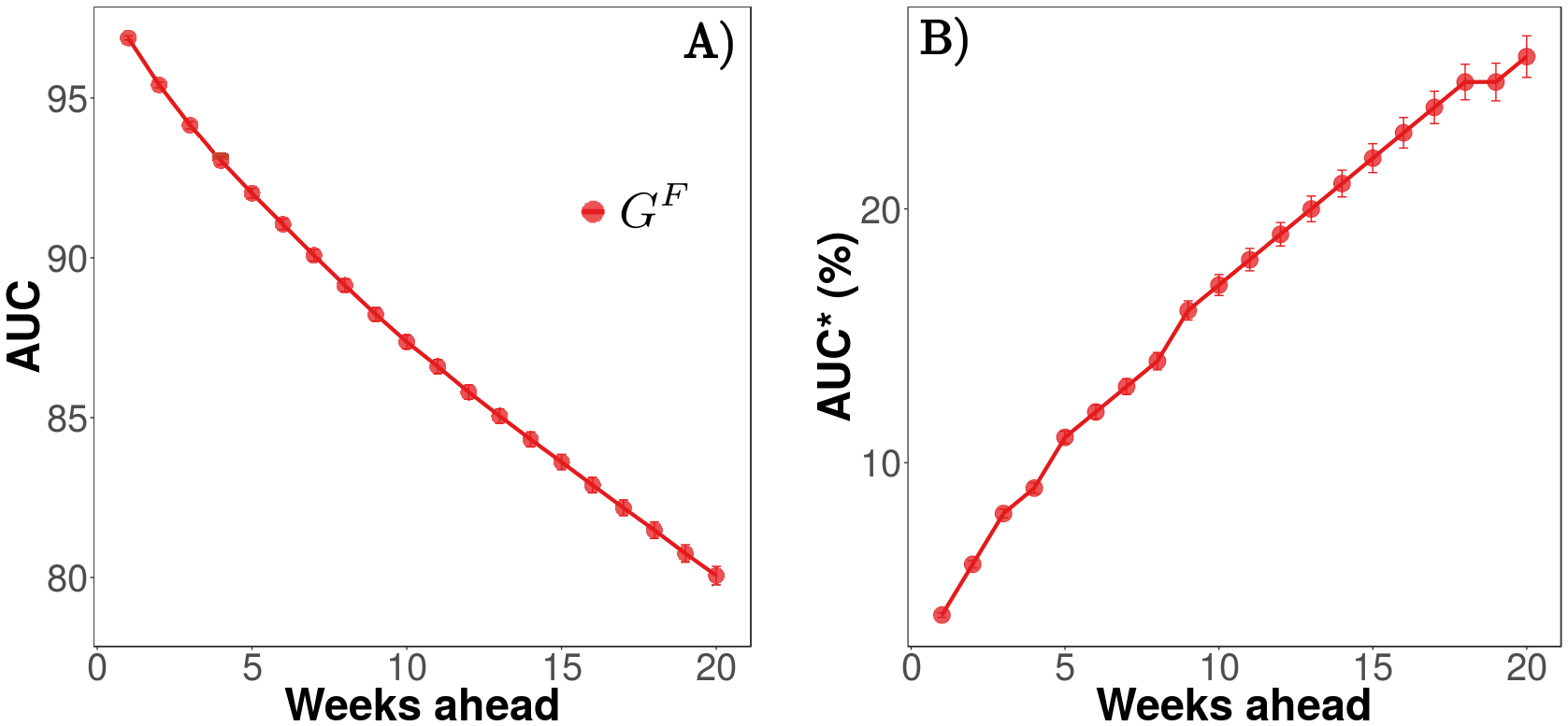}}
\caption{\textbf{The effect of time-lag on out-of-sample predictive performance.} 
Panel A) shows the mean performance (AUC) of the prediction of 
out-of-sample edges of the full financial network $G^{F}$. Panel B) shows the performance improvement ($AUC^*$) against a naive benchmark that assumes that correlation structure is time-invariant, i.e., $G^{F}(t+h) = G^{F}(t)$.
Error bars indicate standard error.
}
\label{fig:TRIADS4-rolling}
 \end{figure}

Fig. \ref{fig:TRIADS4-rolling} reports an aggregate overview of the previous results for the 
out-of-sample prediction in terms of the number of weeks ahead.
We observe that as the lag increases, the prediction performance declines (panel A). 
However, the improvement in performance over the naive benchmark improves (panel B).
 
In Appendix \ref{sec:exp}, we report the results obtained by using an expanding window rather than a rolling window as a training set. 
We observe that expanding the training set does not necessarily lead to better performance. 
In fact, the rolling window analysis yielded better performance overall.

To verify whether the multiplex network provides additional information to that 
from the financial network only, we re-computed the same out-of-sample edge prediction by using the financial network only
and compared this to
the results from the full model that considers both the financial and social information layers.
A comparison between the two models was performed by comparing their respective likelihoods.
We have also disaggregated the prediction of the insertion of new edges $E^+$ and the prediction of edge deletions $E^-$. 
We report the likelihood values and AUC performance obtained for the fit of each model in Table \ref{tb:NA0}.

We observed that the model that includes both financial and social information better fit the data compared to the model that considers financial data only, 
particularly for the case of the prediction of insertion of new edges.
The likelihood ratio increases with prediction lag indicating that full models (i.e. those that consider both financial and social networks) are 
particularly important in long-term link prediction.
Results confirm that the multiplex network is distinctly better than the single financial layer with all likelihood ratios 
having p-value $< 0.001$ for all configurations tested.
\begin{table}[!h]
\centering
\caption{\textbf{Financial Link Prediction Performance Results.} 
High out-of-sample AUCs obtained indicate that the model has high performance balancing 
both false positives and false negatives predictions relative to true positive and negative values.
Log-likelihood ratios ($\lambda$) increase with prediction lag indicating that social media features are particularly important for long-term prediction.
The table reports mean $AUC$ values and log-likelihood ratios $\lambda$ over the test period with corresponding standard deviations in parentheses.
Results are reported for the prediction of new edges $E^*$ and edge deletions $E^-$. 
We also report the average performance $AUC$ obtained in the prediction of the full-graph $G^F$, as well as, 
the performance improvement $AUC^*$ over the benchmark that assumes that correlation structure is time-invariant, i.e., $G^{F}(t+h) = G^{F}(t)$.
Models were trained with a rolling window with initial start and end dates of 09/05/2012 and 09/10/2014, respectively.
The test period ranged from 09/17/2014 to 08/25/2017.
}
\label{tb:NA0}
\resizebox{\textwidth}{!}{%
\begin{tabular}{c|lr|lr|lr}
\multicolumn{1}{c}{} & \multicolumn{2}{c}{$E^+$} & \multicolumn{2}{c}{$E^-$} & \multicolumn{2}{c}{$G^{F}$}\tabularnewline
\cline{2-7}
\multicolumn{1}{l}{Lag}&\multicolumn{1}{c}{$AUC$}&\multicolumn{1}{c}{$\lambda$}&\multicolumn{1}{c}{${AUC}$}&
\multicolumn{1}{c}{$\lambda$}&\multicolumn{1}{c}{$AUC$}&\multicolumn{1}{c}{$AUC^*$ (\%)}\tabularnewline
\hline
$ 1$&87 (0.33)&21 (0.76)&93 (0.11)&34 (1.2)&97 (0.064)&4 (0.091)\tabularnewline
$ 2$&87 (0.37)&33 (1.2)&93 (0.1)&45 (1.5)&95 (0.092)&6 (0.14)\tabularnewline
$ 3$&86 (0.39)&48 (1.5)&93 (0.11)&60 (1.6)&94 (0.11)&8 (0.17)\tabularnewline
$ 4$&86 (0.39)&65 (2)&93 (0.11)&65 (1.9)&93 (0.13)&10 (0.21)\tabularnewline
$ 5$&85 (0.41)&85 (2.6)&93 (0.11)&66 (1.9)&92 (0.15)&11 (0.24)\tabularnewline
$ 6$&85 (0.41)&100 (3.2)&93 (0.1)&74 (2)&91 (0.16)&12 (0.27)\tabularnewline
$ 7$&84 (0.42)&120 (3.5)&93 (0.1)&70 (2.2)&90 (0.18)&13 (0.3)\tabularnewline
$ 8$&84 (0.43)&150 (4.3)&93 (0.1)&72 (1.9)&89 (0.19)&15 (0.33)\tabularnewline
$ 9$&83 (0.44)&180 (5.7)&93 (0.1)&74 (2.2)&88 (0.21)&16 (0.37)\tabularnewline
$10$&83 (0.43)&220 (6.3)&93 (0.096)&79 (1.9)&87 (0.21)&17 (0.4)\tabularnewline
$11$&82 (0.43)&260 (7.2)&93 (0.094)&78 (2)&87 (0.22)&18 (0.43)\tabularnewline
$12$&82 (0.42)&300 (7.9)&93 (0.09)&86 (2.4)&86 (0.22)&19 (0.45)\tabularnewline
$13$&82 (0.43)&330 (7.9)&93 (0.09)&95 (2.1)&85 (0.22)&20 (0.49)\tabularnewline
$14$&81 (0.43)&360 (9.2)&93 (0.084)&100 (2.4)&84 (0.23)&21 (0.51)\tabularnewline
$15$&81 (0.43)&390 (9.9)&93 (0.083)&110 (2.3)&84 (0.24)&22 (0.55)\tabularnewline
$16$&81 (0.43)&410 (10)&93 (0.08)&120 (3)&83 (0.24)&23 (0.58)\tabularnewline
$17$&80 (0.43)&440 (11)&94 (0.079)&130 (2.6)&82 (0.25)&24 (0.62)\tabularnewline
$18$&80 (0.44)&470 (12)&94 (0.076)&150 (3)&82 (0.25)&25 (0.67)\tabularnewline
$19$&80 (0.46)&500 (12)&94 (0.072)&160 (3.6)&81 (0.27)&26 (0.71)\tabularnewline
$20$&80 (0.48)&510 (12)&94 (0.068)&170 (3.7)&80 (0.28)&27 (0.79)\tabularnewline
\hline 
\multicolumn{7}{l}{*A likelihood ratio of $\lambda > 18.47$ indicates statistical significance at $p = 0.001$.}
\end{tabular}
}
\end{table}
 \subsection{Prediction of Social Opinion Structure}
We have so far established that social opinion structure can provide statistically significant information about the future financial market structure.
In this section, we investigate the opposite relationship of whether financial market structure can also significantly improve the prediction 
of future social opinion structure, 
and we determine if this effect is larger or smaller.

\begin{figure}[!h]
\centering
\scalebox{0.38}{\includegraphics{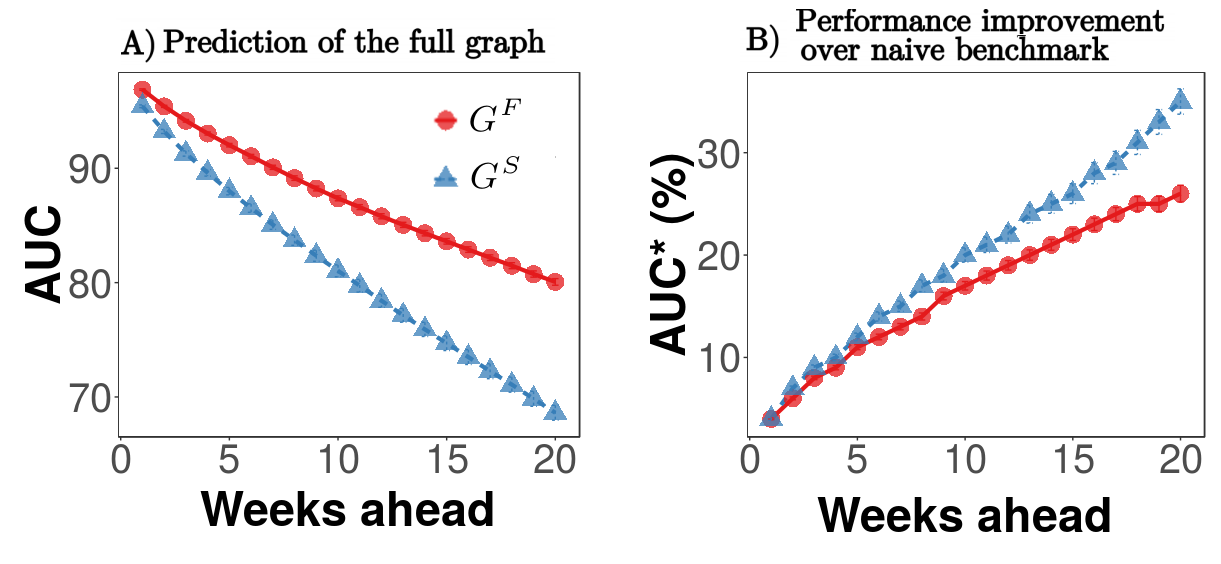}}
\caption{\textbf{Evidence that financial market structure has higher predictability than social media structure.} 
Panel A) shows mean performance (AUC) in the prediction of 
out-of-sample edges of the full financial network $G^{F}$ and the social opinion network $G^{S}$. 
Panel B) shows the performance improvement ($AUC^*$) against a naive benchmark that assumes that the correlation structure is time-invariant.
Error bars indicate standard error.
}
\label{fig:SM-1-rolling}
\end{figure}

The comparison between performance results is summarized in Fig. \ref{fig:SM-1-rolling}, 
where the prediction of social opinion structure $G^{S}$ is plotted together 
with the results for the prediction of financial market structure $G^{F}$ that was discussed previously.
Surprisingly, results suggest that financial market structure has a higher predictability than social opinion structure.
We also observe that both the financial network and social opinion network predictions lead to an improvement 
compared to the naive benchmark that considers time invariance in social network structure.
As previously observed, the relative performance improvement increases with time lag. 
In this case, the relative improvement in prediction is higher for the social opinion structure than for the financial network as observed in Fig. \ref{fig:SM-1-rolling} panel B).


\begin{figure}[!h]
\centering
\scalebox{0.35}{\includegraphics{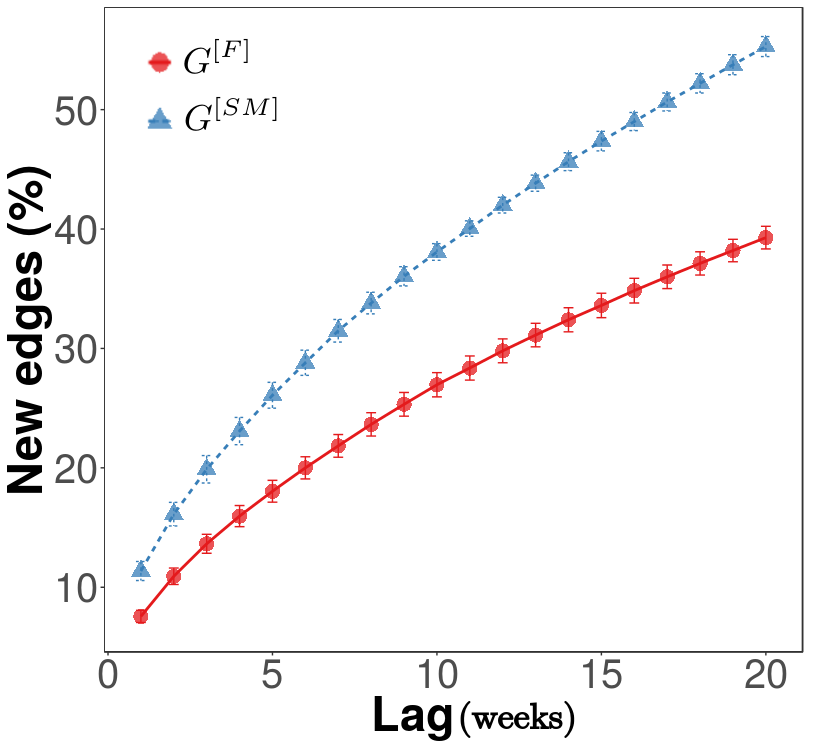}}
\caption{\textbf{Evidence that social media structure is less stable than financial market structure in terms of number of edge changes in time.} We observe that almost 40\% of edges in Financial Networks 
changed after a period of 20 trading weeks while the social media structure changed more than 50\% of its edges over the same time lag.
A network at time $t$ is constructed from a correlation structure estimated from a sliding window of 126 trading days starting at time $t$ that moves with time step of 1 trading week.
The financial network measures co-movement of stock returns while the social network measures co-movement of opinion over the same stocks.
Error bars indicate standard error.
}
\label{fig:SM-2}
\end{figure}

One of the possible reasons why social opinion structure is less predictable compared to financial network structure is the higher structural variability of the former compared to the latter.
Fig. \ref{fig:SM-2} provides evidence that social media structure is less stable than financial market structure in terms of the 
number of edge changes over time.
More edges changed in the social opinion network than in the financial network for all lags tested.
We observed that more than 50\% of the edges in the social media opinion structure 
changed compared to 40\% in the financial network over a time lag of 20 trading weeks.
\section{Discussion and Conclusions}

We investigated whether financial market structure can be better predicted by combining past financial information with past social media sentiment information.
We considered the $N = 100$ most capitalized companies that were part of the S\&P500 index in the period between May 2012 and August 2017.
We generated two networks: A financial network constructed from log-returns of equity prices and a social network constructed from Twitter sentiment analytics. 
We constructed filtered correlation-based networks by keeping the strongest top quartile correlations only that considered a rolling window of $T=126$ trading days.
The two networks were treated as a multiplex problem with two layers of networks that share the same nodes (stocks) but have different edge sets. 

The financial market structure forecasting problem was formulated as a link prediction problem where we estimated the probability
of the addition or removal of a link in the  future on information about the past structure of financial and social opinion networks.

We proposed that financial network links were formed by a combination of the two mechanisms of triadic closure and edge persistence.
The first mechanism assumes that two stocks have a propensity to be correlated if they share common neighbors.
The edge persistence mechanism assumes that two connected stocks tend to remain connected in the future.
A logistic model was trained over a set of data between 09/05/2012 and 09/10/2014 and then results were reported for the 
validation set over the following period from 09/17/2014 and 08/25/2017.

Our results indicate that financial market structure is considerably time variant, 
which invalidates the commonly used assumption of time invariance in the determination of stock correlation structure.
The proposed model exhibited high out-of-sample performance in financial network link prediction, 
particularly in the case of long-term predictions where we observed a performance improvement of up to
40\% over a naive benchmark that assumed that the correlation structure of the financial market was time invariant.
Likelihood ratio analysis demonstrated that models that considered both financial and social information better fit
the data when compared to a restricted model that considers financial information only. 
This provides evidence that supports the use of social information in the prediction of financial market structure.

Finally, our findings indicate that social opinion structure is less stable than financial market structure.
Surprisingly, the prediction of financial market structure using past social and financial information presented higher performance 
compared to the problem of predicting social opinion structure using past social and financial information.

Let us note that network link formation can occur due to mechanisms beyond the ones we studied here.
For instance, networks can form links as a result of a growth process that adds new nodes in the network, e.g., 
IPOs can generate growth in a financial network. 
Among other possible mechanisms, link formation can occur due to preferential attachment, a phenomenon widely observed in real networks where new nodes 
tend to link to the more connected ones \cite{barabasi2016network}.

In summary, this study indicates that social opinion structure is relevant to the prediction of future financial correlation structures.
This result has important consequences because of the fundamental importance of financial correlation structure in Modern Portfolio Theory (MPT) \cite{luenberger2014investment},
Capital Asset Pricing Model (CAPM) and Arbitrage Pricing Theory (APT) \cite{campbell1997econometrics}.
Future work should focus on the investigation of further mechanisms of financial link formation and on applications in portfolio allocation strategies.

\section{Acknowledgments}

This work was supported by PsychSignal.com, which provided social media data. 
T. Aste acknowledges support of the UK Economic and Social Research Council (ESRC) in funding the Systemic Risk Centre (ES/K002309/1). 
T.T.P. Souza acknowledges financial support from the Brazilian National Council for Scientific and Technological Development (CNPq).


\bibliography{biblio}

\begin{thebibliography}{10}
\expandafter\ifx\csname url\endcsname\relax
  \def\url#1{\texttt{#1}}\fi
\expandafter\ifx\csname urlprefix\endcsname\relax\def\urlprefix{URL }\fi
\expandafter\ifx\csname href\endcsname\relax
  \def\href#1#2{#2} \def\path#1{#1}\fi

\bibitem{10.1371/journal.pone.0017994}
M.~Tumminello, S.~Miccichè, F.~Lillo, J.~Piilo, R.~N. Mantegna,
  \href{http://dx.doi.org/10.1371%2Fjournal.pone.0017994}{Statistically
  validated networks in bipartite complex systems}, PLoS ONE 6~(3) (2011)
  1--11.
\newblock \href {http://dx.doi.org/10.1371/journal.pone.0017994}
  {\path{doi:10.1371/journal.pone.0017994}}.
\newline\urlprefix\url{http://dx.doi.org/10.1371%2Fjournal.pone.0017994}

\bibitem{Mantegna1999}
R.~N. Mantegna, \href{http://dx.doi.org/10.1007/s100510050929}{Hierarchical
  structure in financial markets}, The European Physical Journal B - Condensed
  Matter and Complex Systems 11~(1) (1999) 193--197.
\newblock \href {http://dx.doi.org/10.1007/s100510050929}
  {\path{doi:10.1007/s100510050929}}.
\newline\urlprefix\url{http://dx.doi.org/10.1007/s100510050929}

\bibitem{aste2010correlation}
T.~Aste, W.~Shaw, T.~Di~Matteo, Correlation structure and dynamics in volatile
  markets, New Journal of Physics 12~(8) (2010) 085009.

\bibitem{Tumminello201040}
M.~Tumminello, F.~Lillo, R.~N. Mantegna,
  \href{http://www.sciencedirect.com/science/article/pii/S0167268110000077}{Correlation,
  hierarchies, and networks in financial markets}, Journal of Economic Behavior
  \& Organization 75~(1) (2010) 40 -- 58, transdisciplinary Perspectives on
  Economic Complexity.
\newblock \href
  {http://dx.doi.org/http://dx.doi.org/10.1016/j.jebo.2010.01.004}
  {\path{doi:http://dx.doi.org/10.1016/j.jebo.2010.01.004}}.
\newline\urlprefix\url{http://www.sciencedirect.com/science/article/pii/S0167268110000077}

\bibitem{Tumminello26072005}
M.~Tumminello, T.~Aste, T.~Di~Matteo, R.~N. Mantegna,
  \href{http://www.pnas.org/content/102/30/10421.abstract}{A tool for filtering
  information in complex systems}, Proceedings of the National Academy of
  Sciences of the United States of America 102~(30) (2005) 10421--10426.
\newblock \href
  {http://arxiv.org/abs/http://www.pnas.org/content/102/30/10421.full.pdf}
  {\path{arXiv:http://www.pnas.org/content/102/30/10421.full.pdf}}, \href
  {http://dx.doi.org/10.1073/pnas.0500298102}
  {\path{doi:10.1073/pnas.0500298102}}.
\newline\urlprefix\url{http://www.pnas.org/content/102/30/10421.abstract}

\bibitem{1367-2630-12-8-085009}
T.~Aste, W.~Shaw, T.~D. Matteo,
  \href{http://stacks.iop.org/1367-2630/12/i=8/a=085009}{Correlation structure
  and dynamics in volatile markets}, New Journal of Physics 12~(8) (2010)
  085009.
\newline\urlprefix\url{http://stacks.iop.org/1367-2630/12/i=8/a=085009}

\bibitem{song2008analysis}
W.-M. Song, T.~Aste, T.~Di~Matteo, Analysis on filtered correlation graph for
  information extraction, Statistical Mechanics of Molecular Biophysics (2008)
  88.

\bibitem{pozzi2013spread}
F.~Pozzi, T.~Di~Matteo, T.~Aste, Spread of risk across financial markets:
  better to invest in the peripheries, Scientific reports 3.

\bibitem{morales2012dynamical}
R.~Morales, T.~Di~Matteo, R.~Gramatica, T.~Aste, Dynamical generalized hurst
  exponent as a tool to monitor unstable periods in financial time series,
  Physica A: Statistical Mechanics and its Applications 391~(11) (2012)
  3180--3189.

\bibitem{musmeci2014clustering}
N.~Musmeci, T.~Aste, T.~di~Matteo, Clustering and hierarchy of financial
  markets data: advantages of the dbht., CoRR.

\bibitem{song2012hierarchical}
W.-M. Song, T.~Di~Matteo, T.~Aste, Hierarchical information clustering by means
  of topologically embedded graphs, PLoS One 7~(3) (2012) e31929.

\bibitem{curme2015coupled}
C.~Curme, H.~E. Stanley, I.~Vodenska, Coupled network approach to
  predictability of financial market returns and news sentiments, International
  Journal of Theoretical and Applied Finance 18~(07) (2015) 1550043.

\bibitem{1507.00955}
O.~Kolchyna, T.~T.~P. Souza, P.~Treleaven, T.~Aste, Twitter sentiment analysis:
  Lexicon method, machine learning method and their combination, in: G.~Mitra,
  X.~Yu (Eds.), Handbook of Sentiment Analysis in Finance, 2016, Ch.~5.

\bibitem{8252134}
J.~Manfield, D.~Lukacsko, T.~T.~P. Souza, Bull bear balance: A cluster analysis
  of socially informed financial volatility, in: 2017 Computing Conference,
  2017, pp. 421--428.
\newblock \href {http://dx.doi.org/10.1109/SAI.2017.8252134}
  {\path{doi:10.1109/SAI.2017.8252134}}.

\bibitem{1507.00784}
T.~T.~P. Souza, O.~Kolchyna, P.~Treleaven, T.~Aste, \uppercase{t}witter
  sentiment analysis applied to finance: A case study in the retail industry,
  in: G.~Mitra, X.~Yu (Eds.), Handbook of Sentiment Analysis in Finance, 2016,
  Ch.~23.

\bibitem{citeulike:13108056}
I.~Zheludev, R.~Smith, T.~Aste, {When Can Social Media Lead Financial
  Markets?}, Scientific Reports 4.

\bibitem{tetlock2007giving}
P.~C. Tetlock, Giving content to investor sentiment: The role of media in the
  stock market, The Journal of Finance 62~(3) (2007) 1139--1168.

\bibitem{Tobias:2013}
M.~Alanyali, H.~S. Moat, T.~Preis, Quantifying the relationship between
  financial news and the stock market, Sci. Rep. 3.

\bibitem{mao2014quantifying}
H.~Mao, S.~Counts, J.~Bollen, Quantifying the effects of online bullishness on
  international financial markets, European Central Bank Workshop on Using Big
  Data for Forecasting and Statistics, Frankfurt, Germany.

\bibitem{BondtandThaler1985}
W.~F. M.~D. Bondt, R.~Thaler, \href{http://www.jstor.org/stable/2327804}{Does
  the stock market overreact?}, The Journal of Finance 40~(3) (1985) pp.
  793--805.
\newline\urlprefix\url{http://www.jstor.org/stable/2327804}

\bibitem{shleifer2000inefficient}
A.~Shleifer, Inefficient Markets: An Introduction to Behavioral Finance,
  Clarendon Lectures in Economics, OUP Oxford, 2000.

\bibitem{2016arXiv160104535S}
T.~T.~P. {Souza}, T.~{Aste}, {A nonlinear impact: evidences of causal effects
  of social media on market prices}, ArXiv e-prints\href
  {http://arxiv.org/abs/1601.04535} {\path{arXiv:1601.04535}}.

\bibitem{PsychSignal}
\href{https://www.psychsignal.com}{The psychsignal website} (Oct. 2015).
\newline\urlprefix\url{https://www.psychsignal.com}

\bibitem{1402-4896-2003-T106-011}
J.-P. Onnela, A.~Chakraborti, K.~Kaski, J.~Kertész, A.~Kanto,
  \href{http://stacks.iop.org/1402-4896/2003/i=T106/a=011}{Asset trees and
  asset graphs in financial markets}, Physica Scripta 2003~(T106) (2003) 48.
\newline\urlprefix\url{http://stacks.iop.org/1402-4896/2003/i=T106/a=011}

\bibitem{refId0-Onnela-2004}
{Onnela, J.-P.}, {Kaski, K.}, {Kert\'esz, J.},
  \href{https://doi.org/10.1140/epjb/e2004-00128-7}{Clustering and information
  in correlation based financial networks}, Eur. Phys. J. B 38~(2) (2004)
  353--362.
\newblock \href {http://dx.doi.org/10.1140/epjb/e2004-00128-7}
  {\path{doi:10.1140/epjb/e2004-00128-7}}.
\newline\urlprefix\url{https://doi.org/10.1140/epjb/e2004-00128-7}

\bibitem{Battiston2017}
F.~Battiston, V.~Nicosia, V.~Latora,
  \href{http://dx.doi.org/10.1140/epjst/e2016-60274-8}{The new challenges of
  multiplex networks: Measures and models}, The European Physical Journal
  Special Topics 226~(3) (2017) 401--416.
\newblock \href {http://dx.doi.org/10.1140/epjst/e2016-60274-8}
  {\path{doi:10.1140/epjst/e2016-60274-8}}.
\newline\urlprefix\url{http://dx.doi.org/10.1140/epjst/e2016-60274-8}

\bibitem{1367-2630-17-7-073029}
E.~Cozzo, M.~Kivelä, M.~D. Domenico, A.~Solé-Ribalta, A.~Arenas, S.~Gómez,
  M.~A. Porter, Y.~Moreno,
  \href{http://stacks.iop.org/1367-2630/17/i=7/a=073029}{Structure of triadic
  relations in multiplex networks}, New Journal of Physics 17~(7) (2015)
  073029.
\newline\urlprefix\url{http://stacks.iop.org/1367-2630/17/i=7/a=073029}

\bibitem{faraway2006elm}
J.~Faraway, {Extending the Linear Model with R: Generalized Linear, Mixed
  Effects and Nonparametric Regression Models}, CRC Press, 2006.

\bibitem{barabasi2016network}
A.~Barab{\'a}si, M.~P{\~A}3sfai,
  \href{https://books.google.co.uk/books?id=ZVHesgEACAAJ}{Network Science},
  Cambridge University Press, 2016.
\newline\urlprefix\url{https://books.google.co.uk/books?id=ZVHesgEACAAJ}

\bibitem{luenberger2014investment}
D.~Luenberger,
  \href{https://books.google.co.uk/books?id=YMSeDAEACAAJ}{Investment Science},
  Oxford University Press, 2014.
\newline\urlprefix\url{https://books.google.co.uk/books?id=YMSeDAEACAAJ}

\bibitem{campbell1997econometrics}
J.~Campbell, J.~Campbell, A.~Lo, A.~MacKinlay, J.~Champbell, A.~LO,
  A.~MacKinlay, P.~Lo, O.~Campbell,
  \href{https://books.google.co.uk/books?id=lkeKhnqUHx8C}{The Econometrics of
  Financial Markets}, Princeton University Press, 1997.
\newline\urlprefix\url{https://books.google.co.uk/books?id=lkeKhnqUHx8C}

\end{thebibliography}

\section*{Appendix}
\label{sec:support}
\renewcommand\thesection{A\Alph{section}}
\setcounter{section}{0}

\subsection{Ticker Codes of Selected Companies}
\label{sec:comps}
AAPL, AMZN, NFLX, MSFT, GS, GOOGL, BAC, JPM, IBM, DIS, GILD, INTC, YHOO, WMT, GE, XOM, SBUX, CSCO, WFC, NVDA, PCLN, JNJ, MCD, NKE, BA, VZ, ES, 
PFE, KO, CVX, CAT, MU, MRK, CELG, EBAY, MS, CRM, FCX, QCOM, TGT, HD, CHK, BMY, AMGN, PG, HPQ, ORCL, FSLR, WFM, COST, BIIB, PEP, EA, AXP, WYNN, CMCSA, CL, AIG, DOW, NEM, MA, BBY, COP, LOW, TWX, ADBE, HAL, LLY, UNH, LUV, 
MMM, CVS, MO, FDX, DD, ED, KR, MON, UTX, ABT, SLB, YUM, MCO, AMAT, EXPE, AET, DE, GPS, UPS, VLO, CBS, HAS, COH, ALL, WDC, JWN, TXN, PM, UNP, EOG.


\subsection{Prediction Results Using an Expanding Window Training Set}
\label{sec:exp}
In this section, we report results using models that were trained in an expanding window, instead of a rolling window, 
using initial start and end dates of 
09/05/2012 and 09/10/2014, respectively. The test period ranges from 09/17/2014 to 08/25/2017.

  \begin{figure}[!h]
\centering
\scalebox{0.26}{\includegraphics{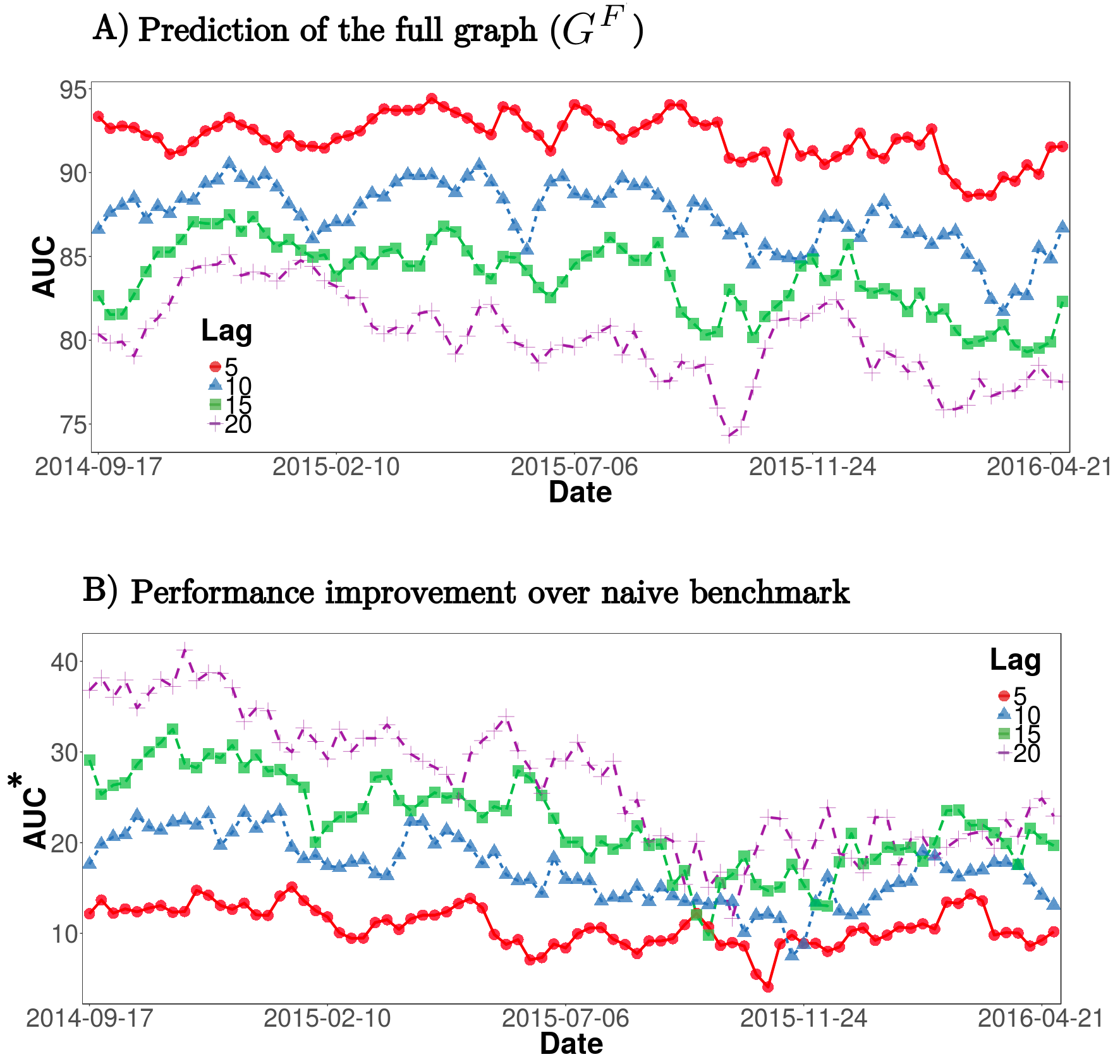}}
\caption{\textbf{Link prediction results using an expanding window training set. 
Evidence of high out-of-sample performance in financial network link prediction.}
Models were trained in an expanding window with initial start and end dates 09/05/2012.
and 09/10/2014, respectively. Test period ranges from 09/17/2014 and 08/25/2017. 
Plots display the performance results (AUC) of a model to predict edges in a financial network at time $t + h$ trained with
information up to date $t$. 
Panel A) shows the performance obtained in the prediction of out-of-sample edges for $h \in (1, 5, 10, 15, 20)$ trading weeks. 
Panel B) shows the performance improvement ($AUC^*$) compared to a naive benchmark that assumes that the 
correlation structure is time-invariant, i.e., $G^{F}(t+h) = G^{F}(t)$.}
\label{fig:TRIADS3}
 \end{figure}
 
  \begin{figure}[!h]
\centering
\scalebox{0.28}{\includegraphics{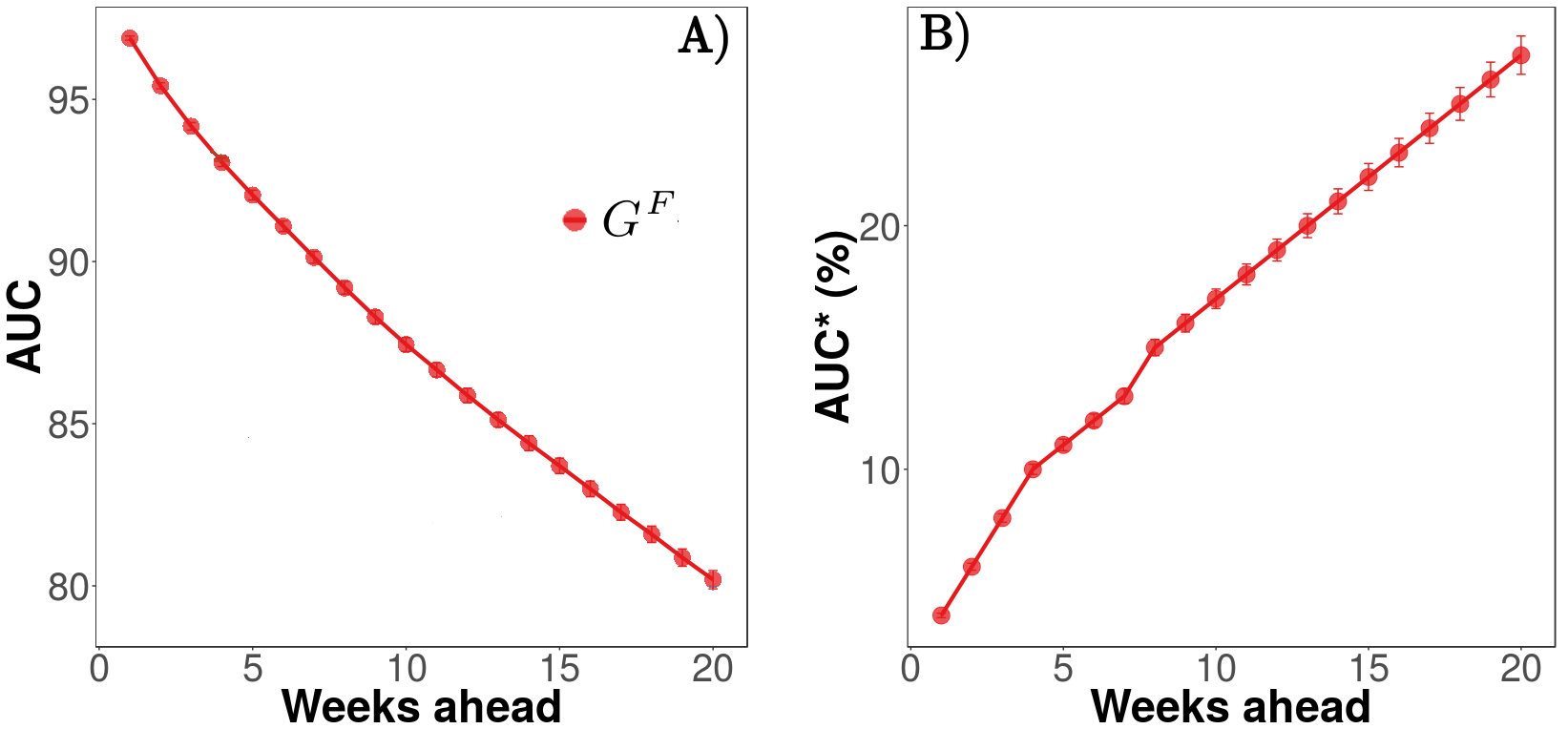}}
\caption{\textbf{Link prediction results using an expanding window training set. The effect of time-lag on out-of-sample predictive performance.}
Panel A) shows the mean performance (AUC) of the prediction of 
out-of-sample edges of the full financial network $G^{F}$. 
Panel B) shows the performance improvement ($AUC^*$) against a naive benchmark that 
assumes that correlation structure is time-invariant, i.e., $G^{F}(t+h) = G^{F}(t)$.
Error bars indicate standard error.
}
\label{fig:TRIADS4}
 \end{figure}

  \begin{figure}[!h]
\centering
\scalebox{0.35}{\includegraphics{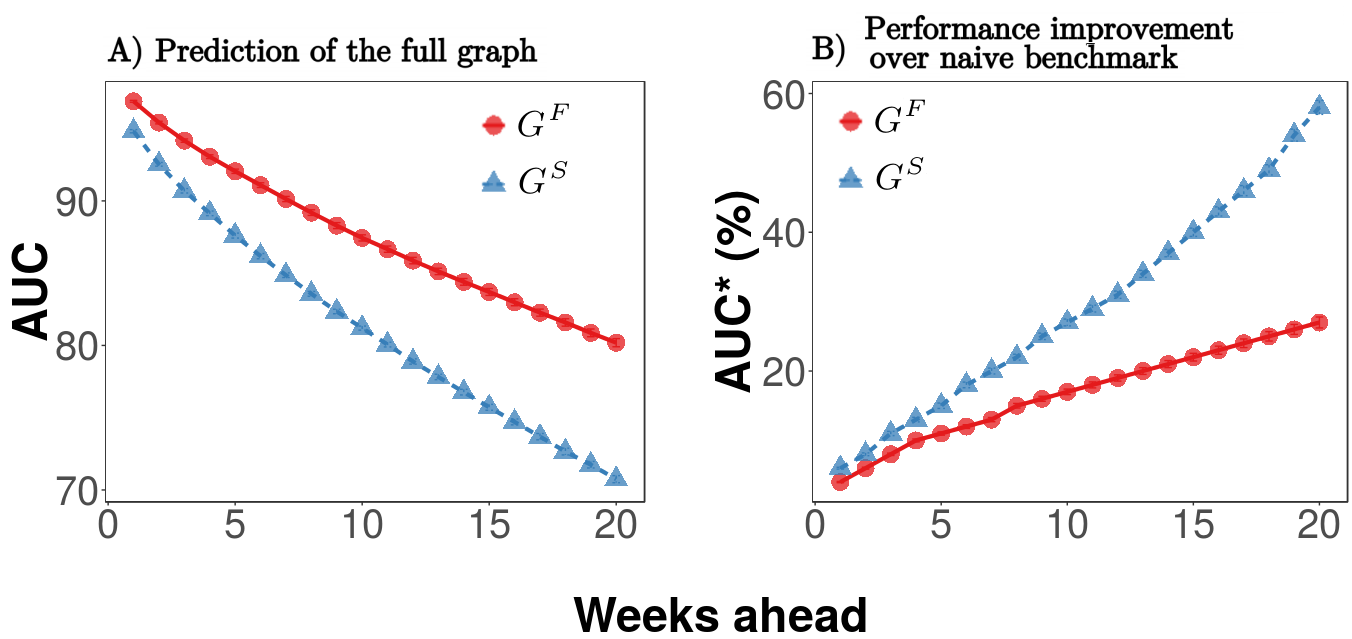}}
\caption{\textbf{Link prediction results using an expanding window training set.
Evidence that social media structure is less stable than financial market structure in terms of number of edge changes in time.} 
We observe that almost 40\% of edges in Financial Networks 
changed after a period of 20 trading weeks while the social media structure changed more than 50\% of its edges over the same time lag.
A network at time $t$ is constructed from a correlation structure estimated from an expanding
window of 126 trading days starting at time $t$ that moves with time step of 1 trading week.
The financial network measures co-movement of stock returns while the social network measures co-movement of opinion over the same stocks.
Error bars indicate standard error.
}
\label{fig:SM-1}
 \end{figure}

\end{document}